\documentclass{XrU2005}
\usepackage{graphicx}

\title{X-Ray Evidence for Multiple Absorbing Structures in Seyfert Nuclei}
\author{J. M. Gelbord}
\affil{MIT Kavli Institute, 77 Massachusetts Ave., Cambridge, MA 02140, USA}
\author{K. A. Weaver}
\affil{NASA GSFC, Code 662, Greenbelt, MD 20771, USA}
\author{T. Yaqoob}
\affil{Department of Physics and Astronomy, Johns Hopkins University,
  3400 N Charles St., Baltimore, MD 21218, USA}

\begin{document}


\maketitle

\begin{abstract}
We have discovered a correlation between the X-ray absorbing column
densities within Seyfert galaxies and the relative alignment between
the central engines and their host galactic disks. This correlation
carries several implications for Seyfert unification models. (1) In
addition to small-scale circumnuclear absorbers, there are absorbing
systems associated with the host galactic plane that are capable of
obscuring the broad line region emission. (2) The misalignment between
the central engine axis and that of the host galaxy arises on
intermediate scales between these absorbers. (3) The small-scale
absorbers have systematically higher column densities and may be
universally Compton-thick.
\end{abstract}

\section{The dual-absorber model}

Seyfert galaxies are generally subdivided into two spectroscopic
classifications:
type 1's have extremely broad permitted emission lines, less broad
forbidden lines, and strong non-thermal continua,
while type 2 Seyferts 
exhibit only the narrow forbidden lines.
Unification models assert that all Seyferts are intrinsically similar
but have different appearances in different directions.  
The canonical
model invokes a parsec-scale torus that hides the innermost, energetic
regions from some lines of sight.  Observers with an
unimpeded view of the central region see a Seyfert 1 and those with
line-of-sight obscuration see a type 2.
To hide the central continuum source and the broad line region (BLR), the
screen must
have $N_{\mathrm{H}} > 10^{21}$~cm$^{-2}$ to attenuate soft X-rays
and be dusty to effectively staunch IR/optical/UV continuum.

The distribution of host galaxy inclinations ($i$) amongst Seyfert types
indicates that a model consisting of a single 
torus is
incomplete.
If such a torus
is the universal source of obscuration,
then one of two scenarios are expected:
either it is aligned with the host galactic plane, causing
Seyfert 1's to be found in face-on hosts and Seyfert 2's in
edge-on galaxies, 
or it is misaligned with the galaxy, in which case there
would be no correlation between Seyfert types and $i$.
Neither is the case.  Several studies have instead established
that type 2 Seyferts are found 
with any $i$ while
type 1's are not found in edge-on galaxies \citep[e.g.,][]{Maiolino95}.
The distribution of Seyfert 2's
suggests an obscuring medium that is misaligned with the
host galaxy, but the dearth of edge-on Seyfert 1 hosts indicates that
there is always sufficient material in the galactic plane to hide the
broad line region.

\begin{figure}
\centering
\includegraphics[width=0.8\linewidth]{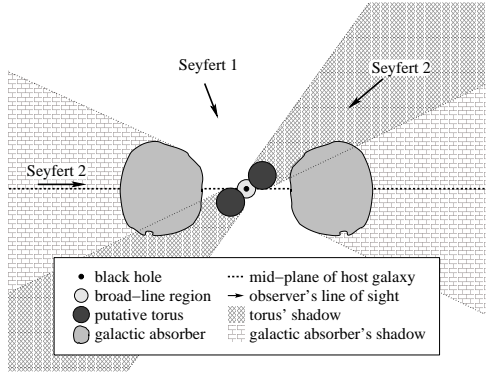}
\caption{The dual absorber framework.  One absorber lies near the
  nucleus and may be identified with the putative torus,
  while the other is on much larger scales and is aligned with the
  host galactic plane.  A type 2 Seyfert is observed if either lies
  along the line of sight to the BLR.\label{fig:model}}
\end{figure}

The distribution of $i$ values can be explained with the introduction of
a second 
absorber
(Fig.~\ref{fig:model}). 
In this model, the small-scale ``nuclear absorber'' or NA (presumably
the torus although other models are possible, e.g., \citealp{Elvis00})
is randomly oriented with respect to the host galaxy;
the second absorber lies at larger scales and is aligned with the
galactic plane, hence the ``galactic absorber'' or GA.
Such a model has been proposed by numerous authors
\citep[e.g.,][]{McLeod95,Maiolino95,Kinney00}.
Several lines of evidence suggest 
an absorbing medium on 100~pc scales, including:
missing edge-on Seyferts of any type from optical, UV and soft X-ray
selected surveys suggesting a large-scale absorber that covers much of
the narrow line region \citep[NLR;][]{McLeod95};
IR reprocessing by dust \citep{Granato97};
and direct imaging of dust lanes on few-hundred pc scales \citep{Malkan98,Pogge02}.
%
%
The relative alignment of the absorbers is an important
parameter of this model.
When the absorbers are misaligned, 
the shadow of the NA covers less of the GA
and the combined covering fraction of the absorbers
increases. 
Seyfert 1's should tend to be in well-aligned systems and Seyfert 2's
attenuated by the GA and not the NA should prefer poorly-aligned ones.

The two absorbers should differ in their mean column densities.
A significant fraction of Seyfert 2 galaxies exhibit Compton thick
absorption ($N_{\mathrm{H}} > 10^{24}$~cm$^{-2}$).
To provide marginally 
Compton-thick absorption over a covering fraction $f$ requires
$10^9 M_{\odot} f (r_{\mathrm{100~pc}})^2$; 
a reasonable quantity for the NA but an excessive amount for the GA.
Dynamical mass measurements of some nearby Seyferts can
rule out any appreciable Compton-thick covering fraction at $\sim$100~pc
scales \citep[e.g.,][]{Maiolino98}.
Thus, most if not all Compton thick Seyferts are attenuated by their NA,
and typical lines of sight through the GA
will have much lower column densities.

To test the dual-absorber model we combine measurements of the line
of sight attenuation with geometric constraints on the internal
alignment.
We divide the Seyferts into three classes:
unobscured (optically-defined Seyfert 1's), 
modestly obscured (Compton-thin Seyfert 2's), 
and
heavily obscured (Compton-thick or nearly so),
with the latter two differentiated by X-ray spectroscopy.
We assume that 
these respectively correspond to lines of sight that are unobstructed,
intercept only the GA, and intercept the NA. 
Rather than model $N_{\mathrm{H}}$ values, we rely upon the equivalent
width (EW) of the Fe K$\alpha$ line 
to avoid an ambiguity of models fitted to low S/N data.
When the continuum around 6~keV is repressed 
(if $N_{\mathrm{H}} > 10^{23.5}$~cm$^{-2}$)
the EW of the 6.4~keV Fe line skyrockets, providing a robust
indicator of heavy obscuration.
As an alignment measure,
we use published values of
$\delta$, the angle between the radio jet and the host galaxy major
axis \citep{Kinney00}:
misaligned systems have low values and perfectly aligned ones have
$\delta = 90^\circ$.

\section{Results from the ASCA Sample}

\begin{figure}
\centering
\rotatebox{270}{\includegraphics[height=0.8\linewidth]{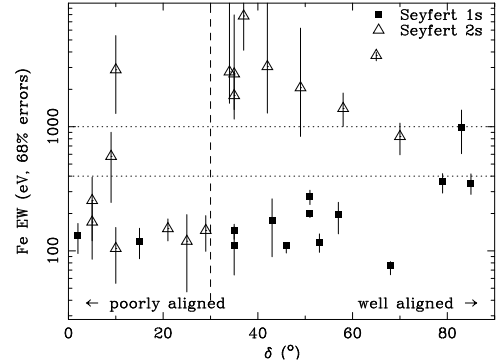}}
\caption{The distribution of the alignment parameter $\delta$ amongst the
  Seyfert classes.  Heavily obscured sources are defined as those with
  Fe EW $>$ 1~keV and modestly obscured sources have EW $<$ 400~eV.
  The preferences of low-EW Seyfert 2's for low $\delta$ values and
  Seyfert 1's for high values is predicted by the dual-absorber
  model.\label{fig:feEWdelta}}
\end{figure}

We analyzed all 31 Seyferts with ASCA detections and published $\delta$
values, classifying each as heavily, modestly, or not obscured
(Fig.~\ref{fig:feEWdelta}).
We find that 
(1) modestly obscured Seyfert 2's all have $\delta < 30^{\circ}$;
(2) unobscured systems prefer moderate-to-high $\delta$ values
($\delta > 30^{\circ}$);
(3) heavily obscured systems have no strong correlation with $\delta$;
and (4) when taken together, systems with modest or no obscuration are
uncorrelated with $\delta$.
%
These distributions all agree with the dual-absorber model (points 3
\& 4 \textit{should} be independent of $\delta$ because they depend
only upon whether our line of sight intercepts the NA).
A KS test shows the $\delta$ distributions of unobscured and modestly
obscured Seyferts to differ with 99.8\% confidence.
The strength of the correlation of modestly obscured sources and $\delta$
is surprising; we would expect some misaligned systems to have
low $\delta$ values due to projection effects.

These observations allow us to make some inferences about Seyfert
structures.
If the misalignment must be severe before we see GA-only
attenuation, then the GA must have a much smaller covering fraction
than the NA (contrary to Fig.~\ref{fig:model}).
The fact that we have any correlations with $\delta$ means that the
radio jet is a reliable indicator of the direction of the NA.
Thus, the misalignment between the central engine and the host galaxy
must take place on intermediate scales.
The NA seldom if ever has a column density below $10^{23}$~cm$^{-2}$.
Otherwise some fraction of modestly absorbed systems would be observed
through the NA and hence shouldn't correlate with $\delta$. 
The strength of the correlation 
argues against this but needs to be tested with a larger, more
complete sample.

\bibliographystyle{XrU2005}
\bibliography{refs.bib}

\end{document}